\documentclass[12pt]{article}
\usepackage[cp1251]{inputenc}
\usepackage[english]{babel}
\usepackage{amsfonts}

\begin{document}

\input epsf
\def\rot{\operatorname{rot}}

\begin{center}
{\Large
Real Commuting Differential Operators Connected
with~Two-Dimensional Abelian Varieties

}
\end{center}
\begin{center}
{A.E. Mironov}
\end{center}

 \section{Introduction}

In~[1] A.~Nakayashiki constructed commutative rings of
($g!\times g!$)-matrix partial differential operators in~
$g$
variables (see also~[2]).
The common vector eigenfunctions and eigenvalues
of these operators are parametrized by points
of some principally polarized abelian  variety of dimension
$g$
with a~nonsingular theta-divisor.
Each operator corresponds to some meromorphic function
(spectral function)
on this abelian  variety with a~pole on the theta-divisor.
Henceforth these operators are referred to as
{\it Nakayashiki operators}.

 Some explicit formulas  for Nakayashiki operators with
$g=2$ were obtained in~[3].
 Using these formulas, we find smooth real operators.

Theorems~1 and~2 are the main results of this article.

{\bf Theorem~1.} {\sl
For
$g=2$,
there are no Nakayashiki operators with smooth real
doubly periodic coefficients
but there are Nakayashiki operators with real singular
doubly periodic coefficients.}

This theorem is an~analog of the theorem
of Feldman, Knorrer, and Trubowitz~[4] who demonstrated that
a~two-dimensional Schr\"odinger operator without magnetic field
which has a~smooth doubly periodic real potential can be
finite-gap only at one energy level; i.e.,
the Bloch functions (the eigenfunctions of the Schr\"odinger
operator
and the  translation operators by periods)
may be parametrized by a~Riemann surface of finite genus
only at one energy level.
Theorem~1 means that there is no smooth real
Nakayashiki operators that are finite-gap at all energy levels.
Nevertheless, there exist real Nakayashiki operators
with singular coefficients which  are finite-gap at every energy
level.

We take as the abelian  variety  the Jacobi variety
of a~Riemann surface of genus~2 with real branching points.
In this case the symmetric matrix
$\Omega$
of the periods of basis abelian  differentials
has purely imaginary components~[5]. Introduce
the magnetic translation operators~
$T_1^*$
and~
$T_2^*$:
$$
T_1^*\varphi(y)=\varphi(y+e_1)\exp(2\pi y_1),\quad
T_2^*\varphi(y)=\varphi(y+e_2)\exp(2\pi y_2),
$$
where
$y=(y_1,y_2),$
$e_j$
is the $j$th row of the imaginary part of the period matrix~
$\Omega$.
The magnetic translation operators differ from the translation
operators only by an~exponential twist.
The arguments of exponential functions in
the magnetic translation operators are chosen so that
$A_i(y+e_j)-A_i(y)=2\pi \delta_{ij}$,
where
$(A_1,A_2)$
is a~vector-potential of the magnetic field [6]. Then the operators
$T_j^*$
commute with the  covariant differentiation operators
$\partial_{y_i}-A_i$.
The operators
$T_1^*$
and
$T_2^*$
commute with one another. This is a~consequence of the fact that
in our case the magnetic flux through the elementary cell formed
by the vectors~
$e_1$ and~$e_2$ vanishes.
In the general case we have
$T_1^*T_2^*=T_2^*T_1^*\exp(ie\Phi)$, with
$e$
the charge and
$\Phi$
the magnetic flux~ [6], and the operators~
$T_1^*$
and
$T_2^*$
commute if
$\frac{e\Phi}{2\pi}$
is integral.

An~eigenfunction of a~matrix differential operator
is called a~{\it magneto-Bloch function\/} if its components are
eigenfunctions of the magnetic translation operators.
Denote the theta-function of the abelian  variety
${\Bbb C}^2/\{{\Bbb Z}^2+\Omega {\Bbb Z}^2\}$
by
$\theta(z),$
with
$z=(z_1,z_2)$.

{\bf Theorem~2.} {\sl
There exist Nakayashiki operators
with smooth real coefficients.
The diagonal of the operator
$H$
corresponding to the function
$
\partial_{z_1}^2\ln\theta(z)+
\partial_{z_2}^2\ln\theta(z)
$
is composed of Schr\"odinger operators of the form
$$
H_{11}=(\partial_{y_1}-A_1)^2+
(\partial_{y_2}-A_2)^2+u(y),
\quad
H_{22}=(\partial_{y_1}-\widetilde{A}_1)^2+
(\partial_{y_2}-\widetilde{A}_2)^2+\tilde{u}(y)
$$
with doubly periodic magnetic fields
${\rm rot} (A_1,A_2,0)$
and
${\rm rot} (\widetilde{A}_1,\widetilde{A}_2,0)$
and with doubly periodic potentials
$$
u(y+e_j)=u(y),
\quad
\tilde{u}(y+e_j)=\tilde{u}(y).
$$
The components of the vector-potentials satisfy the equalities
$$
A_i(y+e_j)-A_i(y)=
\widetilde{A}_i(y+e_j)-
\widetilde{A}_i(y)=2\pi \delta_{ij}.
$$
The magneto-Bloch functions of
$H$
at each energy level are parametrized by Riemann surfaces
of finite genus.
The components of ~
$H$
commute with
$T_1^*$
and
$T_2^*.$}

We also indicate Nakayashiki operators of simplest form.
For example, the operators
$L$
and
$L_1$
corresponding to the functions
$\partial_{z_1}^2\log\theta(z)$
and
$\partial_{z_1}\partial_{z_2}\log\theta(z)$
are as follows:

{\bf Lemma 1.} {\sl
The following hold:
$$
L=
\left(
\begin{array}{cc}
-\partial^2_{x_1}+c_1\partial_{x_2}+U & W \\
 \frac{V}{c_2}(-\partial^2_{x_1}+c_1\partial_{x_2}+U-c_3) &
 -\partial^2_{x_1}-c_1\partial_{x_2}+
   \widetilde{U}+\frac{WV}{c_2}
\end{array}\right),
$$
where
$$
U=\partial_{x_1}^2\ln V+
(\partial_{x_1}\ln V+c_4)^2-
c_1(\partial_{x_2}\ln V+c_5)+c_3,
$$
$$
\widetilde{U}=\partial_{x_1}^2\ln W+
(\partial_{x_1}\ln W-c_4)^2+
c_1(\partial_{x_2}\ln W-c_5)+c_3;
$$
and
$$
L_1=
\left(
\begin{array}{cc}
-\partial_{x_1}\partial_{x_2}+
U_2\partial_{x_1}+c_6\partial_{x_2}+U_1 & W_1 \\
 \frac{V_1}{c_2}(-\partial^2_{x_1}+c_1\partial_{x_2}+U-c_3) &
-\partial_{x_1}\partial_{x_2}+
\widetilde{U}_2\partial_{x_1}-c_6\partial_{x_2}+
\widetilde{U}_1+\frac{WV_1}{c_2}
\end{array}\right),
$$
where
$$
U_1=\partial_{x_1}\partial_{x_2}\ln V+
(\partial_{x_1}\ln V+c_4)(\partial_{x_2}\ln V+c_5)-
U_2(\partial_{x_1}\ln V+c_4)-
c_6\partial_{x_2}\ln V+c_7,
$$
$$
\widetilde{U}_1=\partial_{x_1}\partial_{x_2}\ln W+
(\partial_{x_1}\ln W-c_4)(\partial_{x_2}\ln W-c_5)-
\widetilde{U}_2(\partial_{x_1}\ln W-c_4)+
c_6\partial_{x_2}\ln W+c_7,
$$
$$
W_1=\frac{c_6}{c_1}W-\frac{1}{2c_1}\partial_{x_1}W,
\quad
V_1=\frac{c_6}{c_1}V+\frac{1}{2c_1}\partial_{x_1}V,
$$
$$
U_2=\frac{1}{2c_1}(U+c_8),
\quad
\widetilde{U}_2=-\frac{1}{2c_1}(\widetilde{U}+c_8),
$$
and
$c_j$
are some constants $($see ~{\rm (15)--(18)).}
}

Observe that the coefficients of~
$\partial_{x_2}$
in the 11- and 22-components of these operators are
constants and the coefficients of~
$\partial_{x_1}$
in ~
$L$
vanish. Moreover, all coefficients of ~
$L$
and~
$L_1$
are rationally expressed in terms of
$V$
and
$W$
and their derivatives. There are no similar relations between the
coefficients
of the operators in~[3].

We indicate partial solutions to the system of nonlinear equations
$[L,L_1]=0$
in~
$V$
and~
$W$.
The solutions are given by~(8) and~(9).

{\bf Theorem~3.} {\sl
The coefficients of the Nakayashiki operators
are rationally expressible in terms of the coefficients~
$V$
and~
$W$
of ~
$L$
and their derivatives.
}

Observe that the 11-components of the operators commute
modulo a~heat operator (Lemma~8);
i.e., for arbitrary two 11-components~
$A$
and~
$B$
there is an~operator
$C$
such that
$$
[A,B]=C\bigl(-\partial^2_{x_1}+c_1\partial_{x_2}+U-c_3\bigr).
$$

The coefficients of the Nakayashiki operators cannot
satisfy evolution equations like
the Kadomtsev--Petviashvili hierarchy (KP)
$$
[\partial_{t_n}-L_n,\partial_{t_m}-L_m]=0.
$$
Indeed, to the space variables there correspond
$g$
linearly independent rectilinear windings on
the $g$-dimensional torus. To the time variables there also
correspond
rectilinear windings on the torus; moreover, they are linear
combinations
of spatial windings, since the dimension of the torus
coincides with the number of the space variables.
Consequently, the time derivatives of the coefficients of operators
can be expressed linearly in terms of the
derivatives with respect to the space variables and,
changing variables, we can reduce ``evolution'' equations
to   commutation equations for Nakayashiki operators.
In the particular case of~
$g=1$
the finite-gap solutions of the KP hierarchy
(the so-called stationary solutions)
are not interesting, since replacement of differentiation
with respect to time by differentiation with respect to the space
variables
takes the equations of the hierarchy into the commutation equations
$[L_n,L_m]=0$ for the operators.

Let us show that  Nakayashiki's construction~[1]
does not lead to evolution equations.
Below all notations untill Section~2 are taken from~[1].
Introduce two functions
$$
\varphi_1=\frac{\theta(z+(c'-x'd'-x_0,x'))}{\theta(z)}\exp,
$$
$$
\varphi_2=\frac{\theta(z+(c'-x'd'-x_0,x')+c'')\theta(z-c'')}
{\theta^2(z)}\exp,
\quad
 c''\in{\Bbb C}^2,c''\not\in{\Bbb Z}^2+\Omega{\Bbb Z}^2,
$$
$$
\exp=\exp\biggl(\,\sum_{i=0}^1\sum_{n\geq \delta_{i0}}t_{n,(i)}
\frac{(-1)^n}{n!}(u_{n,(i)}(z)+d_i(1-\delta_{i0})u_{n+1,(0)}(z))
\biggr),
$$
where
$x'$
and
$x_0$
are the space variables,
$u_{n,(i)}$
are the  derivatives of the logarithm of the theta-function,
$c'$, $d_i$, and~$\delta_{i,0}$
are constants,
$x_0=t_{1,(0)}$, and $x'=t_{0,(1)}$.
These functions determine a~basis for the free module
$B_{ct}$
over~
${\cal D}_t$,
where
${\cal D}_t$
is the ring of differential operators in the variables
$x_0$
and~
$x'$
with analytic coefficients
depending on~$t_{n,(i)}$ in a~neighborhood of~$0$.
In~[1]  some embedding $\iota$ was constructed
of the
${\cal D}_t$-module
$B_{ct}$
into the ring of pseudodifferential operators.
Equations~(6.8) of~[1] read:
$$
\frac{\partial W_i}{\partial t_{\beta}}+
W_i\partial^{\beta}=
\sum_{j=1}^{2}B_{i,\beta,j}W_j,
$$
where
$W_i=\iota(\varphi_i)$
and
$B_{i,\beta,j}\in {\cal D}_t$.
Observe that the image of
$\varphi\in B_{ct}$
under
$\iota$
depends only on
$\frac{\varphi}{\exp}$.
Consequently, the operators
$W_i$
are independent of time, and
$\frac{\partial W_i}{\partial t_{\beta}}=0.$
Hence, ~(6.8) of~[1] are not evolution equations.

In Section~2 we recall Nakayashiki's construction and prove
Theorems~1 and~2.

In Section~3 we prove Lemma~1 and Theorem~3.

The author is grateful to I.~A. Ta\u\i manov
for posing the problem as well as for useful discussions and remarks.

\section{ Smooth Real Operators}

We start with recalling Nakayashiki's construction
of the Baker--Akhiezer module~
$M_c$.
Then we prove Theorem~1.
Next we introduce a~hyperelliptic surface~
$\Gamma$
of genus~2 with real branching points and
take a~canonical basis of cycles on it by utilizing the scheme
of~[5].
Thereafter we introduce four real two-dimensional tori
$T_j$,\ $1\leq j\leq 4$,
in
$X$
(the Jacobi variety of~
$\Gamma$).
The theta-function takes real values on these tori.
In Lemma~3 we prove that the theta-function has no zeros on ~
$T_1$.
In Lemma~4 we find the intersection points of the theta-divisor
and the  translated theta-divisor.
The proof of Theorem~2 consists in verifying
that the coefficients of the Nakayashiki operators of the second and
third
orders from Propositions~1--3 of~[3], with changes made below,
are real and smooth.

Consider the principally polarized complex abelian  variety
$X={\Bbb C}^2/\{{\Bbb Z}^2+\Omega {\Bbb Z}^2\}$, where
$\Omega$
is a~symmetric
$2\times 2$-matrix with
${\rm Im}\Omega>0$.
The theta-function is defined by the series
$$
\theta(z)=\sum_{n\in {\Bbb Z}^2}\exp
(\pi i \langle \Omega n, n \rangle+2\pi i
\langle n, z\rangle ),\quad  z\in {\Bbb C}^2,
$$
where
$\langle n, z\rangle=n_1z_1+n_2z_2.$
It has the periodicity properties
$$
\theta (z+\Omega m+n)=\exp(
-\pi i \langle \Omega m, m\rangle -2\pi i
\langle m,z \rangle)\theta (z), \quad  m, n\in {\Bbb Z}^2.
$$
Let
${\cal D}$
be the ring
${\cal O}[\partial_{x_1},\partial_{x_2}]$ of differential operators,
where
${\cal O}$
is the ring of analytic functions
(of~$x$)
in a~neighborhood of~
$0\in{\Bbb C}^2$.
In~[1] Nakayashiki introduced the Baker--Akhiezer module~
$M_c$
over
${\cal D}$
which consists of functions of the form
$$
f(z,x)\exp(-x_1\partial_{z_1}\log\theta(z)-x_2\partial_{z_2}
\log\theta(z)).
$$
The function
$f(z,x)$
is meromorphic on~
${\Bbb C}^2\times U_f$,
where
$U_f$
is some neighborhood of~~
$0\in{\Bbb C}^2$,
has a~pole in the theta-divisor
$\Theta$
(the zeros of the theta-function
$\theta(z)$),
and possesses the periodicity property
$$
f(z+\Omega m+n,x)=\exp(-2\pi i\langle m,c+x\rangle)f(z,x),
\quad
m,n\in {\Bbb Z}^2,
\quad
c=(c_1,c_2)\in{\Bbb C}^2.
$$

The following was proven in~[1]:

{\bf Nakayashiki's Theorem.} {\sl
If
$\Theta$
is a~nonsingular variety and
$c\ne 0$
then
$M_c$
is a~free
${\cal D}$-module of rank~
$2$.
}

Fix a~basis
$\Phi_c=(\phi_{1c}(z,x),\phi_{2c}(z,x))^{\top}$
for the ${\cal D}$-module
$ M_c$.
Denote by
${\cal A}_{\Theta}$
the ring of meromorphic functions on~
$X$
with a~pole in
$\Theta$.
Take
$\lambda (z)\in
{\cal A}_\Theta$.
Since
$M_c$
is a~free ${\cal D}$-module,
there is a~unique
($2\times 2$)-matrix operator
$L_{\Phi_c}(\lambda )$
with components in~
${\cal D}$
such that
$$
L_{\Phi_c}(\lambda)\Phi_c=\lambda\Phi_c, \eqno{(1)}
$$
where
$\lambda\Phi_c=(\lambda\phi_{1c},\lambda\phi_{2c})^{\top}$.
Since
$L_{\Phi_c}(\lambda)$
are differential operators in the variables
$x_j$
while
$\lambda$
depends only on
$z$,
(1) implies the commutation condition
$$
L_{\Phi_c}(\lambda\mu)=L_{\Phi_c}(\lambda)L_{\Phi_c}(\mu)=
L_{\Phi_c}(\mu)L_{\Phi_c}(\lambda),
$$
where $\mu(z)\in {\cal A}_{\Theta}$.
We thus arrive at the following

{\bf Corollary \rm [1].} {\sl
There is an~embedding of the rings
$$
L_{\Phi_c}:{\cal A}_{\Theta}\rightarrow {\rm {Mat}}(2,{\cal D}),
$$
where ${\rm {Mat}}(2,{\cal D})$
is the ring of $(2\times 2)$-matrix differential operators.
The range of the embedding is a~commutative ring
of differential operators.
}

Now, we suppose that
$\Theta$
is a~nonsingular Riemann surface.

{\sc Proof of Theorem~1.}
We start with proving the second part of the theorem. Introduce the
following
functions in~
$M_c$:
$$
\psi=\frac{\theta(z+c+x)}{\theta(z)}
\exp(-x_1\partial_{z_1}\log\theta(z)
-x_2\partial_{z_2}\log\theta(z)),
$$
$$
\psi_{c'}=\frac{\theta(z+c+c'+x)\theta(z-c')}
{\theta^2(z)}
\exp(-x_1\partial_{z_1}\log\theta(z)
-x_2\partial_{z_2}\log\theta(z)).
$$
They determine a~basis for the ${\cal D}$-module~
$M_c$~[3].
In~[3] we have found the Nakayashiki operators
$L_{c,c'}({\cal A}_{\Theta})$
in this basis. The coefficients of the operators~
$L_{c,c'}(\lambda)$
are doubly periodic; moreover,
the coefficients of all components of each operator have
singularity at
$x=-c$~
[3].
Let
$X$
be the Jacobi variety of a~Riemann surface with real branching
points.
In this case we have
$\bar{\theta}(z)=\theta(\bar{z})$
~[5].
Suppose also that
$c,c'\in{\Bbb R}^2$
and
$\lambda(z)$
is a~real function for
$z\in{\Bbb R}^2$
(for example,
$
\lambda=\partial_{z_j}\partial_{z_i}
\log\theta(z)
$).
The fact that the operators
$L_{c,c'}(\lambda)$ are real
is immediate from the fact that the eigenfunction
$
(\psi,\psi_{c'})^{\top}
$
of the operator
$L_{c,c'}(\lambda)$
is real for real values of ~
$z$.

We turn to proving the first part of the theorem.
First of all, observe that
$M_0$ is not a~free ${\cal D}$-module of rank~2.
This follows, for example, from the fact that
$\partial_{x_j}\psi(z,0)=0$ which
implies that a~function of the form
$
f(z,x)\exp(-x_1\partial_{z_1}\log\theta(z)
-x_2\partial_{z_2}\log\theta(z))\in M_0,
$
where
$f(z,x)$
has a~pole of the second order on the theta-divisor,
cannot in general be represented  as
$d_1\psi+d_2\psi_{c'}$
for
$x=0$, where
$d_1\in{\cal D}$
is a~first-order operator and
$d_2$
is the operator of multiplication by a~function; however,
this is possible for
$x\ne 0$.

Let
$X$
be the Jacobi variety of the Riemann surface~
$\Gamma$;
$L(\lambda)$,
a~Nakayashiki operator with doubly periodic
coefficients with periods
$\tau_1,\tau_2\in{\Bbb R}$;
$\Psi=(\psi_1,\psi_2)^{\top}$,
a~Bloch eigenfunction whose components
are  basis elements of the ${\cal D}$-module~
$M_c$;
and
$$
L(\lambda)\Psi=\lambda\Psi,\quad
 \Psi(z,x_1+\tau_1,x_2)=
-\partial_{z_1}\log\theta(z)\tau_1\mu_1
\Psi(z,x_1,x_2),
$$
$$
\Psi(z,x_1,x_2+\tau_2)=
-\partial_{z_2}\log\theta(z)\tau_2\mu_2\Psi(z,x_1,x_2),
$$
where
$\mu_1,\mu_2\in{\Bbb C}.$
Suppose that
$L(\lambda)$
has real coefficients for
$x\in{\Bbb R}^2.$
Then
$X$
admits an~antiholomorphic involution
$\tau$
such that
$$
\overline{\Psi}(z,x)=
\Psi(\tau(z),x),
\quad
\bar{\lambda}(z)=\lambda(\tau(z)).
$$
Moreover,
$\tau$
leaves the theta-divisor invariant,
$\tau(\Theta)=\Theta$,
since the multiplicative functions
$\partial_{z_1}\log\theta(z)$
and
$\partial_{z_2}\log\theta(z)$
have a~pole on the theta-divisor.
Consequently,
$\tau$
is induced by an~antiholomorphic involution of~
$\Gamma$
and coincides with complex conjugation.
The theta-divisor is invariant under
$\tau$
only if~
$\Gamma$
has real branching points~[5].

Let us show that
$c\in{\Bbb R}^2.$
Denote
the transition operator   from the basis
$\psi,\psi_{c'}$
to the basis
$\psi_1,\psi_2$ by ~$A$.
The equalities
$\Psi=A(\psi,\psi_{c'})^{\top},$
$\bar{\theta}(z)=\theta(\bar{z})$,
and
$\Psi(z+\Omega m,x)=\exp(-2\pi i\langle m$, $c\rangle)\Psi(z,x)$ then
imply
that
$\overline{\Psi}(\overline{z+\Omega m},x)=
\exp(-2\pi i\langle m, \bar{c}\rangle)
\overline{\Psi}(\bar{z},x)$.
Hence, from
$\Psi(z,x)=\overline{\Psi}(\bar{z},x)$
we obtain
$c\in{\Bbb R}^2.$

The cause for nonsmoothness of the operators is as follows.
Replace
$x+c$
with
$x$
and divide
$\psi_1$
and
$\psi_2$
by
$
\exp(-c_1\partial_{z_1}\log\theta(z)-
c_2\partial_{z_2}\log\theta(z))$
to obtain the functions
$\widetilde{\psi}_1$, $\widetilde{\psi}_2\in M_0.$
Since the ${\cal D}$-module~
$M_0$
is not free of rank~2, the equality
$\widetilde{L}(\lambda)
\widetilde{\Psi}=\lambda\widetilde{\Psi}$,
$\widetilde{\Psi}=(\widetilde{\psi}_1,\widetilde{\psi}_2)^{\top}$,
does not hold in a~neighborhood of~
$x=0$.
Consequently, the coefficients of
$\widetilde{L}(\lambda)$
have singularity at
$x=0$, and so
the coefficients of
$L(\lambda)$
have singularity at
$x=-c$.
Namely, if the coefficients of ~
$A$
and~
$A^{-1}$
are smooth at
$x=-c$
then the equality~
$A^{-1}L(\lambda)A=L_{c,c'}(\lambda)$ [3]
implies that the operator
$L(\lambda)$
is nonsmooth at
$x=-c$.
If ~
$A$
or
$A^{-1}$
is nonsmooth at
$x=-c$
then nonsmoothness of
$L(\lambda)$
results from the equality
$L(\lambda)=AL_{c,c'}(\lambda)A^{-1}$
and the fact that all components of the operator~
$L_{c,c'}(\lambda)$
are nonsmooth for
$x=-c$.
Theorem~1 is proven.

Denote by
$\Gamma$
the smooth completion of the Riemann surface that is defined
in the $(y,w)$-plane by the equation
$w^2=P(y)=(y-y_1)\dots (y-y_5)$
with real
$y_1<\dots <y_5$.
Denote by
$X$
the Jacobi variety of the hyperelliptic surface~
$\Gamma$.
The holomorphic involution
$\sigma:(y,w)\rightarrow (y,-w)$ acts on~$\Gamma$
with the fixed points
$Q_i=(y_i,0),i=1,\dots,5, \infty$,
and we also have the antiholomorphic involution
$\tau:(y,w)\rightarrow(\bar{y},\bar{w})$.
The involution
$\tau$
has three fixed cycles:
$$
C_1:\{y_1\leq y\leq y_2, w=\pm\sqrt{P(y)}\},
\quad
C_2:\{y_3\leq y\leq y_4, w=\pm\sqrt{P(y)}\},
$$
$$
C_3:\{y_5\leq y\leq \infty, w=\pm\sqrt{P(y)}\}.
$$
Take a~canonical basis
$a_1$, $a_2$, $b_1$, $b_2$
of cycles with the intersection indices
$a_i\circ a_j=b_i\circ b_j=0$ and $a_i\circ b_j=\delta_{ij}$
as shown in Fig.~1
(the dots indicate parts of cycles
on the ``lower leaf'' of the Riemann surface).
The union of the cycles~
$C_1$, $C_2$, and
$C_3$
divides ~
$\Gamma$
into two disjoint pieces.


Observe that
$a_1=C_1$ and $a_2=C_2$.
The antiholomorphic and holomorphic involutions act
on these cycles as follows:
$$
\tau a_1=a_1,\quad  \tau a_2=a_2,\quad  \tau b_1=-b_1,\quad  \tau
b_2=-b_2,
$$
$$
\sigma a_1=-a_1,\quad  \sigma a_2=-a_2,\quad  \sigma b_1=-b_1,\quad
\sigma b_2=-b_2,
$$
where equality is understood to be in the  homology group.
 On the corresponding canonical basis of abelian
differentials~
$\omega_1$
and~
$\omega_2$
such that
$$
\int\limits_{a_j}\omega_i=\delta_{ij},
\quad
\int\limits_{b_j}\omega_i=\int\limits_{b_i}\omega_j=\Omega_{ij},
\quad i,j=1,2,
$$
the involutions
$\tau$
and~
$\sigma$ act as follows:
$$
\tau^*\omega_i=\overline{\omega}_i,\quad
\sigma^*\omega_i=-\omega_i.
                                \eqno{(2)}
$$
Introduce four real tori in
$X$:
$$
T_1: z\equiv i(t_1,t_2),\quad
T_2: z\equiv i(t_1,t_2)+\left(\frac{1}{2},0\right),
$$
$$
T_3:z\equiv i(t_1,t_2)+\left(0,\frac{1}{2}\right),\quad
T_4:z\equiv i(t_1,t_2)+\left(\frac{1}{2},\frac{1}{2}\right),
$$
where
$(t_1,t_2)\in{\Bbb R}^2$,
the symbol
$\equiv$
stands for equality modulo an~element of the lattice
${\Bbb Z}^2+\Omega{\Bbb Z}^2$, and
the matrix
$\Omega$
is composed of the entries
$\Omega_{ij}.$
The theta-function is real on these tori.

{\bf Lemma 2.} {\sl
The following equalities are valid:
$$
\int\limits_{\infty}^{Q_1}\omega
\equiv
\left(\frac{\Omega_{11}}{2},\frac{\Omega_{12}}{2}\right),
\quad
\int\limits_{\infty}^{Q_2}\omega\equiv
\left(\frac{1}{2}+\frac{\Omega_{11}}{2}, \frac{\Omega_{12}}{2}
\right),
$$
$$
\int\limits_{\infty}^{Q_3}\omega\equiv
\left(\frac{1}{2}-\frac{\Omega_{21}}{2},
-\frac{\Omega_{22}}{2}\right),\quad
\int\limits_{\infty}^{Q_4}\omega\equiv
\left(\frac{1}{2}-\frac{\Omega_{21}}{2},
\frac{1}{2}-\frac{\Omega_{22}}{2}\right),
\quad
\int\limits_{\infty}^{Q_5}\omega\equiv
\left(\frac{1}{2},\frac{1}{2}\right).
$$
}

{\sc Proof} Join
$\infty$
and
$Q_1$
by an~oriented path
$l$
such that
$l\cup -\sigma l=b_1$
($-\sigma l$
stands for~
$\sigma l$
with opposite orientation). Then
$$
\int\limits_l\omega-\int\limits_{\sigma l}\omega=
\int\limits_{b_1}\omega,
\quad
\int\limits_l\omega=\int\limits_{\sigma l}\sigma^*\omega=
-\int\limits_{\sigma l}\omega;
$$
consequently,
$$
\int\limits_{\infty}^{Q_1}\omega
=\frac{1}{2}\int\limits_{b_1}\omega\equiv
\left(\frac{\Omega_{11}}{2},\frac{\Omega_{12}}{2}\right).
$$
Similarly, we demonstrate that
$$
\int\limits_{Q_1}^{Q_2}\omega\equiv
\left(\frac{1}{2},0\right),
\quad
\int\limits_{Q_2}^{Q_3}\omega=\frac{1}{2}\int\limits_{b_1-b_2}\omega
\equiv
\left(\frac{\Omega_{11}}{2}-\frac{\Omega_{21}}{2},
\frac{\Omega_{12}}{2}-
\frac{\Omega_{22}}{2}\right),
$$
$$
\int\limits_{Q_3}^{Q_4}\omega=\frac{1}{2}\int\limits_{a_2}\omega
\equiv
\left(0,\frac{1}{2}\right),
\quad
\int\limits_{Q_4}^{Q_5}\omega=\frac{1}{2}\int\limits_{b_2}\omega
\equiv
\left(\frac{\Omega_{21}}{2},\frac{\Omega_{22}}{2}\right).
$$
The lemma is proven.

\newpage
\begin{picture}(170,100)(-100,-80)
\qbezier(0,0)(0,50)(75,50)
\qbezier(75,50)(150,50)(150,0)
\qbezier[30](150,0)(150,-50)(75,-50)
\qbezier[30](75,-50)(0,-50)(0,0)

\qbezier(-17,0)(-17,18)(7,18)
\qbezier(7,18)(31,18)(31,0)
\qbezier(31,0)(31,-18)(7,-18)
\qbezier(7,-18)(-17,-18)(-17,0)

\qbezier(50,0)(50,15)(75,15)
\qbezier(75,15)(100,15)(100,0)
\qbezier(100,0)(100,-15)(75,-15)
\qbezier(75,-15)(50,-15)(50,0)

\qbezier(80,0)(80,15)(105,15)
\qbezier(105,15)(130,15)(130,0)
\qbezier[12](130,0)(130,-15)(105,-15)
\qbezier[12](105,-15)(80,-15)(80,0)

\put(-8,0){\line(2,0){25}}
\put(60,0){\line(2,0){30}}
\put(118,0){\line(2,0){65}}

\put(-10,-9){\shortstack{$y_1$}}
\put(15,-9){\shortstack{$y_2$}}
\put(58,-9){\shortstack{$y_3$}}
\put(85,-8){\shortstack{$y_4$}}
\put(113,-9){\shortstack{$y_5$}}
\put(180,-9){\shortstack{$\infty$}}
\put(15,20){\shortstack{$a_1$}}
\put(75,55){\shortstack{$b_1$}}
\put(75,20){\shortstack{$a_2$}}
\put(105,20){\shortstack{$b_2$}}
\put(60,-70){\shortstack{Fig. 1.}}

\put(-8,0){\circle{1}}
\put(17,0){\circle{1}}
\put(60,0){\circle{1}}
\put(90,0){\circle{1}}
\put(118,0){\circle{1}}
\put(183,0){\circle{1}}

\put(10,18){\vector(-1,0){6}}
\put(75,15){\vector(-1,0){6}}
\put(75,50){\vector(-1,0){6}}
\put(106,15){\vector(-1,0){6}}
\end{picture}

As shown in~[7],
with this choice of a~canonical basis of cycles and the choice of
$\infty$
as the initial point of the Abel mapping,
the vector of Riemann constants equals~
$$
K\equiv
\left(\frac{\Omega_{11}}{2}+\frac{\Omega_{12}}{2},
\frac{\Omega_{21}}{2}+
\frac{\Omega_{22}}{2}\right)+\left(1,\frac{1}{2}\right).
$$

{\bf Lemma 3.} {\sl
The theta-function
$\theta(z)$
has no zeros on ~
$T_1$.
}

{\sc Proof.}
Suppose that
$z$
belongs to
$T_1$
and to the theta-divisor. Then
$\bar{z}\equiv -z$
and, by the Riemann theorem about the  zeros of a~theta-function
(see~[8]),
$z\equiv A(P)+K,$
where
$A(P)$
is the Abel mapping with the initial point
$\infty$.
From (2) and the fact that
$\Omega$
is a~purely imaginary matrix we obtain the equalities
$$
\bar{z}\equiv
A(\tau(P))-
\left(\frac{\Omega_{11}}{2}+\frac{\Omega_{12}}{2},
\frac{\Omega_{21}}{2}+\frac{\Omega_{22}}{2}\right)+
\left(1,\frac{1}{2}\right)
\\
\equiv
-z\equiv
$$
$$
-A(P)-
\left(\frac{\Omega_{11}}{2}+\frac{\Omega_{12}}{2},
\frac{\Omega_{21}}{2}+\frac{\Omega_{22}}{2}\right)-
\left(1,\frac{1}{2}\right);
$$
consequently,
$A(\tau(P))+A(P)\equiv 0$.
Together with (2), this yields
$\tau(P)=\sigma(P)$.
Hence, either
$P$
is a~branching point or the $y$-coordinate of~
$P$
is real and the $w$-coordinate of~$P$ is purely imaginary.
These points constitute the three cycles
$$
B_1:\{\infty\leq y\leq y_1,\  w=\pm\sqrt{P(y)}\},
\quad
B_2:\{y_2\leq y\leq y_3,\  w=\pm\sqrt{P(y)}\},
$$
$$
B_3:\{y_4\leq y\leq y_5,\  w=\pm\sqrt{P(y)}\}.
$$
For the points of these cycles we have
$$
\overline{A(P)}\equiv A(\tau(P))\equiv A(\sigma(P))\equiv -A(P).
$$
Hence,
$$
A(P)\equiv
i(t_1,t_2)+\frac{(m,n)}{2},\quad  (t_1,t_2)\in {\Bbb R}^2,\
m,n\in {\Bbb Z}.
$$
Consequently, the real part of~
$A(P)$
does not change upon the circuit around the cycles
$B_j$, $j=1,2,3.$
Since
$Q_1\in B_1$, $Q_2\in B_2$, and $Q_5\in B_3$,
Lemma~2 implies the inclusions
$A(B_1)\subset T_1$, $A(B_2)\subset T_2$, and $A(B_3)\subset T_4$.
Therefore,
$z\equiv A(P)+K$
cannot belong to~
$T_1$.
The lemma is proven.

Put
$$
c'\equiv
\left(\frac{\Omega_{11}}{2}-\frac{\Omega_{21}}{2},
\frac{\Omega_{12}}{2}-\frac{\Omega_{22}}{2}\right).
$$

{\bf Lemma 4.} {\sl
The theta-divisor and the Riemann surface defined in~
$X$
by the equation
$\theta(z-c')=0$
intersect
at the two points
$$
p_1
\equiv
\left(\frac{\Omega_{12}}{2}+\frac{1}{2},
\frac{\Omega_{22}}{2}+\frac{1}{2}\right),
\quad
p_2
\equiv\left(\frac{\Omega_{11}}{2}+\frac{1}{2},
\frac{\Omega_{21}}{2}+\frac{1}{2}\right).
$$
}

{\sc Proof.}
A~point
$z$
belongs to the theta-divisor whenever it has the form
$z\equiv \int\nolimits_{\infty}^{P}\omega+K,$
where
$P\in\Gamma$.
Consequently, the intersection points look like
$$
p_1\equiv\int\limits_{\infty}^{P_1}\omega+K,
\quad
p_2\equiv\int\limits_{\infty}^{P_2}\omega+K,
$$
where
$P_1$
and
$P_2$
are zeros of the function
$\theta(\int\nolimits_{\infty}^P\omega+K-c')$
on~
$\Gamma$.
The function
$\theta(\int\nolimits_{\infty}^P\omega+K-c')$
is not identically zero on~
$\Gamma$
(since
$K-c'\equiv(1,\frac{1}{2})\not\equiv K$).
Then by the Riemann theorem
$$
\int\limits_{\infty}^{P_1}\omega+\int\limits_{\infty}^{P_2}\omega
\equiv c'
$$
and
the points
$P_1$
and
$P_2$
are determined uniquely by~
$c'$ (see, for example,~[7]).
By Lemma~2,
$$
\int\limits_{\infty}^{Q_2}\omega+\int\limits_{\infty}^{Q_3}\omega
\equiv c';
$$
consequently,
$$
p_1\equiv\int\limits_{\infty}^{Q_2}\omega+K,
\quad
p_2\equiv\int\limits_{\infty}^{Q_3}\omega+K.
$$
The lemma is proven.

From now on, we assume that
$c\in T_1$.
Henceforth we use the following notations:
$$
\theta_j(z)=\partial_{z_j}\theta(z),\quad
\theta_{kj}(z)=\partial_{z_k}\partial_{z_j}\theta(z),\quad
k,j=1,2.
$$

{\sc Proof of Theorem~2.}
To find smooth real operators, we must make
some modifications in the formulas of~[3]
where, proving Propositions~2 and~3, we used the fact that
$\theta(\Delta)=0$,
where
$\Delta$
stands for the vector of Riemann constants. The other properties of
$\Delta$
are not used in the proof of these propositions.
All formulas for the Nakayashiki operators
of Propositions~2 and~3 of~[3] remain valid if we replace
$\Delta$
with
$p_3=\int\nolimits_{\infty}^{P_3}\omega+K$,
where
$P_3$
is an~arbitrary point of
$B_2$
other than~
$Q_2$
or
$Q_3$ and
$\theta(p_3)=0,\ p_3\in T_4.$
 Calculating
$[L_{c,c'}(\partial_{z_k}\partial_{z_j}\log\theta(z))]_{12}$
in Proposition~2 of~[3],
we put
$z=0$.
The formula remains valid if we put
$z=p_4$,
where
$p_4$
is an~arbitrary point of~
$T_4$
such that
$\theta(p_4)\ne 0.$
Then
$$
[L_{c,c'}(\partial_{z_k}\partial_{z_j}\log\theta(z))]_{12}=
\frac{\theta^2(p_4)}{\theta(p_4+c+c'+x)\theta(c')}
$$
$$
\times\bigl(\partial_{z_k}\partial_{z_j}-
f_{c,c'}^{kj}\partial_{z_1}-g_{c,c'}^{kj}\partial_{z_2}
-h_{c,c'}^{kj}
+2\partial_{z_k}\partial_{z_j}
\log\theta(z)\bigr)
\left(\frac{\theta(z+c+x)}{\theta(z)}\right)
\biggr\vert_{z=p_4}.
$$
Similarly, in the formula for
$g^j_{c,c'}$
of Proposition~3
we now put
$z=p_4$
instead of~
$z=0$.
It is these formulas that we will use below.

The constants
$\theta_k(p_j)$
are imaginary and the functions
$\partial_{x_k}\log\theta(p_j+c+x)$
are purely imaginary for
$x\in i{\Bbb R}^2+(\frac{1}{2},\frac{1}{2})$;
consequently, the functions
$f^{kj}_{c,c'}$
and
$g^{kj}_{c,c'}$
in Proposition~2 of~[3] are purely imaginary and
$h^{kj}_{c,c'}$
are real. Hence, the operators
$[L_{c,c'}(\partial_{z_j}\partial_{z_k}\log\theta(z))]_{11}$
and
$[L_{c,c'}
(\partial_{z_j}\partial_{z_k}\log\theta(z))]_{12}$
are real. The numbers
$\alpha_{kj}$
and
$\alpha$
in Proposition~2 of~[3] are real, since the functions
$\partial_{z_j}\partial_{z_k}\log\theta(z)$
and
$\frac{\theta(z-c')\theta(z+c')}{\theta^2(z)}$
are real on
$T_1$
and the functions
$\partial_{z_j}\partial_{z_k}\log\theta(z)$
are linearly independent
(this follows, for example, from Nakayashiki's theorem, since
the 11-coefficients of the operators
$L_{c,c'}(\partial_{z_j}\partial_{z_k}\log\theta(z))$
are equal to
$\partial_{x_j}\partial_{x_k}$).
Consequently, the operators
$$
[L_{c,c'}(\partial_{z_j}\partial_{z_k}\log\theta(z))]_{21},
\quad
[L_{c,c'}(\partial_{z_j}\partial_{z_k}\log\theta(z))]_{22}
$$
are real.

Since the coefficients of the Nakayashiki operators
are expressed in terms of the theta-function and its derivatives
while
the theta-function is real-valued on
$T_1$
and
$T_4$,
the fact that the coefficients of the operators are real for
$x\in i{\Bbb R}^2$
implies that so are the coefficients for
$x\in i{\Bbb R}^2+(\frac{1}{2},\frac{1}{2})$.
From Proposition~3 of~[3] we infer that the coefficients of the
 operators~
$Z_1$
and
$Z_2$
[2] are purely imaginary for
$x\in i{\Bbb R}^2$.
Then Proposition~1 implies that the operators
$L_{c,c'}(i\partial_{z_s}\partial_{z_j}\partial_{z_k}\log\theta(z))$
are real for
$x\in i{\Bbb R}^2$.

 By
Propositions~1,~2, and 3 of~[3]
with the above-made modifications, to prove smoothness
we have to demonstrate that the functions
$\theta(p_1+c+x)$, $\theta(p_2+c+x)$, $\theta(p_3+c+c'+x)$,
and
$\theta(p_4+c'+c+x)$
do not vanish for
$x\in i{\Bbb R}^2+(\frac{1}{2},\frac{1}{2}).$
This follows from Lemma~3.

Introduce the magnetic translation operators
$\widetilde{T}_1$
and~
$\widetilde{T}_2$:
$$
\widetilde{T}_1\varphi(x)=\varphi(x+
\Omega_1)\exp(2\pi i x_1),
\quad
\widetilde{T}_2\varphi(x)=\varphi(x+\widetilde{\Omega}_2)
\exp(2\pi i x_2),
$$
where
$\Omega_j$
is the $j$th row of
$\Omega.$
The functions
$\psi$
and
$\psi_{c'}$
are magneto-Bloch functions:
$$
\widetilde{T}_1\psi=\mu_1\psi,\quad
\mu_1=\exp(-\pi i \Omega_{11}-2\pi i (z_1+c_1)-
\Omega_{11}\partial_{z_1}\log\theta(z)-
\Omega_{12}\partial_{z_2}\log\theta(z)),
$$
$$
\widetilde{T}_2\psi=\mu_2\psi,\quad
\mu_2=\exp(-\pi i \Omega_{22}-2\pi i (z_2+c_2)-
\Omega_{12}\partial_{z_1}\log\theta(z)-
\Omega_{22}\partial_{z_2}\log\theta(z)),
$$
$$
\widetilde{T}_1\psi_{c'}=\mu_1\exp(-2\pi i c_1')\psi_{c'},
\quad
\widetilde{T}_2\psi_{c'}=\mu_2\exp(-2\pi i c'_2)\psi_{c'}.
$$
Instead of
$\psi_{c'}$
we use the function
$$
\widetilde{\psi}_{c'}=\psi_{c'}
\exp\left(2\pi i\left\langle x-\left(\frac{1}{2},\frac{1}{2}\right),
\Omega^{-1}c'\right\rangle\right).
$$
Then by symmetry of~$\Omega$
$$
\widetilde{T}_1\widetilde{\psi}_{c'}=\mu_1
\widetilde{\psi}_{c'},
\quad
\widetilde{T}_2\tilde{\psi}_{c'}=\mu_2
\widetilde{\psi}_{c'}.
$$
The Nakayashiki operators in the basis
$\psi$, $\widetilde{\psi}_{c'}$
look like
$dL_{c,c'}(\lambda)d^{-1}$,
where
$d$
is the diagonal matrix with diagonal
$(1, \exp(2\pi i\langle x-(\frac{1}{2},\frac{1}{2})$,
$\Omega^{-1}c'\rangle))$
[3], and are smooth and real under the same conditions as the
operators
$L_{c,c'}(\lambda)$.
Denote by
$H$
the smooth operator
$dL_{c,c'}(\partial_{z_1}^2\log\theta(z)
+\partial_{z_2}^2\log\theta(z))
d^{-1}$
which is real for
$x\in i{\Bbb R}^2+(\frac{1}{2},\frac{1}{2})$.
Its 11-component is
$$
H_{11}=(i\partial_{x_1}-A_1)^2+
(i\partial_{x_2}-A_2)^2+u(x),
$$
where [3]
$$
A_1=\frac{i}{2}\bigl(f^{11}_{c,c'}+f^{22}_{c,c'}\bigr),
\quad
A_2=\frac{i}{2}\bigl(g^{11}_{c,c'}+g^{22}_{c,c'}\bigr) .
$$
This is the Schr\"odinger operator in the periodic magnetic field
${rm rot} (A_1,A_2,0)$.
For the components of the vector-potential
$(A_1,A_2)$
we have the equality
$$
A_k(x+\Omega_j)-A_k(x)=2\pi\delta_{kj}.
$$
The magnetic translation operators commute with the
covariant differentiation operators:
$$
\widetilde{T}_j(i\partial_{x_k}-A_k)=
(i\partial_{x_k}-A_k)\widetilde{T}_j.
$$
The magneto-Bloch function
$\psi$
for
$z\in\Gamma_{c'}$
satisfies the Schr\"odinger equation with the Hamiltonian~
$H_{11}$
and the energy
$\partial_{z_1}^2\log\theta(z)+\partial_{z_2}^2\log\theta(z).$
Hence, the potential is doubly periodic; i.e.,
$u(x+\Omega_1)=u(x+\Omega_2)=u(x)$,
and the Schr\"odinger operator commutes
with the magnetic translation operators:
$\widetilde{T}_jH_{11}=H_{11}\widetilde{T}_j.$
The 12-component of
$H$
is the operator of multiplication by a~doubly periodic function. The
 operator
$H_{21}$
looks like
$F(x)\widetilde{H}_{21}$,
where
$F$
is a~doubly periodic function and
$\widetilde{H}_{21}$
is a~second-order operator with constant leading coefficients.
The 22-component is as follows:
$$
H_{22}=(i\partial_{x_1}-\widetilde{A}_1)^2+
(i\partial_{x_2}-\widetilde{A}_2)^2+\tilde{u}(x).
$$
The operator
$H_{22}$
possesses the same properties as
$H_{11}.$
In particular,
$H_{22}$
commutes with the magnetic translation operators.

The magneto-Bloch functions of
$H$
at an energy level
$\lambda$
are parametrized by the Riemann surface defined in
$X$
by the equation
$$
\partial_{z_1}^2\log\theta(z)+\partial_{z_2}^2\log\theta(z)=\lambda.
$$
To complete the proof, we have to change coordinates
and recall that
$\partial_{x_k}$
are the operators of complex differentiation; i.e.,
$\partial_{x_k}=\frac{1}{2}(\frac{\partial}
{\partial \tilde{x}_k}-
i\frac{\partial}{\partial y_k}).$

Theorem~2 is proven.

\section{ The Nakayashiki Operators}

In the beginning of this section we introduce two Riemann surfaces
$\Gamma_1$
and
$\Gamma_2$
of genus~2 embedded in the two-dimensional abelian  variety
$X$.
In Lemma~5, we use the Fay formula~(3) to prove that
$\Gamma_1$
and
$\Gamma_2$
are tangent to the theta-divisor. In~(5) and~(6), we indicate a~basis
$\psi_1$, $\psi_2$
for the Baker--Akhiezer module. It follows from Lemma~5 that
$\psi_1$
and
$\psi_2$,
bounded on~
$\Gamma_1$
and~
$\Gamma_2$ respectively,
are one-point Baker--Akhiezer functions~[9].
In Lemma~7 we find some coefficients
of the 11- and 12-components of the second-order operators.
In Lemma~8 we indicate a~connection between the 11- and 12-components
and between the 21- and 22-components of the operators.
In Lemma~9 we find some coefficients of ~
$L_1$.
Lemma~1 ensues from Lemmas~7--9.
In Lemma~10 we prove that the coefficients of the 11-components
of the Nakayashiki operators are rationally expressible in terms of
the function~
$V$
and its derivatives.
Theorem~3 ensues from Lemmas~11 and~12. In the latter
we prove that the coefficients
of the operators of the second and third orders
are rationally expressible in terms of ~
$V$,
~$W$,
and their derivatives.

The abelian  variety
$X$
is the Jacobi variety of some Riemann surface~
$\Gamma$
of genus~2.
There is a~canonical basis
$a_1$, $a_2$, $b_1$, $b_2$
of cycles on~
$\Gamma$
with the intersection indices
$a_i\circ a_j=b_i\circ b_j=0$
and
$a_i\circ b_j=\delta_{ij}$,
and there is a~basis of abelian  differentials~
$\omega_1$
and~
$\omega_2$
such that
$\int\nolimits_{a_j}\omega_i=\delta_{ij},i,j=1,2$, and
the components of ~
$\Omega$
are equal to
$\Omega_{ij}=\int\nolimits_{b_j}\omega_i$.
The following identity due to J.~D. Fay~[10]
is valid for points $\widetilde{R},\widetilde{Q}\in\Gamma$:
$$
\sum_{i,j=1}^2F_iG_j\partial_{z_i}\partial_{z_j}
\log\theta(z)=\tilde{c}_3+
\tilde{c}_2
\frac
{\theta\bigg(z+\int\limits_{\widetilde{Q}}^{\widetilde{R}}\omega
\bigg)
\theta\bigg(z+\int\limits_{\widetilde{R}}^{\widetilde{Q}}
\omega\bigg)}
{\theta^2(z)},
$$
where
$F_i=\frac{\omega_i(\widetilde{R})}{dr},
G_i=\frac{\omega_i(\widetilde{Q})}{dq}$,
$r$ and $q$
are local parameters in neighborhoods of ~
$\widetilde{R}$
and~
$\widetilde{Q}$,
$$
 \int\limits_{\widetilde{Q}}^{\widetilde{R}}\omega=
\biggl(\,\int\limits_{\widetilde{Q}}^{\widetilde{R}}\omega_1,
\int\limits_{\widetilde{Q}}^{\widetilde{R}}\omega_2\biggr),
$$
and
$\widetilde{c_2}$ and $\widetilde{c_3}$
are some constants. Denote by
$R$, $Q\in\Gamma$
the zeros of~
$\omega_2$.
Since  $\int\nolimits_Q^R\omega=-2K,$ where
$K$
is the vector of Riemann constants with respect to~
$Q$
(see, for example, [11]), we have
$$
\partial^2_{z_1}\log\theta(z)=c_3+
c_2\frac{\theta(z-2K)\theta(z+2K)}
{\theta^2(z)},
                                \eqno{(3)}
$$
where
$c_2$
and
$c_3$
are some constants (we write them down explicitly in~(15)).
Denote by
$\Gamma_1$
and
$\Gamma_2$
the Riemann surfaces that are defined in~
$X$
by the equations
$\theta(z+2K)=0$
and
$\theta(z-2K)=0$.

{\bf Lemma 5.} {\sl
The Riemann surfaces
$\Theta$
and
$\Gamma_1$
$(\Theta$
and
$\Gamma_2)$
intersect at the point~
$-K$
($K$)
with multiplicity~$2$ {\rm(}are tangent$)$, and
$\theta_{11}(-K)=\theta_{11}(K)\ne 0$.
}

{\sc Proof.}
The  index of intersection of~
$\Theta$
and
$\Gamma_1$
equals~2 (see~[8]). Suppose that
$\Theta$
and
$\Gamma_1$
intersect at two different points.
Let
$A:\Gamma\rightarrow X$
be the Abel mapping given by the formula
$$
A(P)=\biggl(\,\int\limits_{Q}^P\omega_1,\int\limits_{Q}^P\omega_2
\biggr),
\quad P\in\Gamma.
$$
By the Riemann theorem about the  zeros of a~theta-function, the
equality
$\theta(z)=0$
amounts to the fact that
$z=A(P)+K.$
Consequently,
$z=A(R)+K=-K$
is the intersection point of~
$\Theta$
and
$\Gamma_1$.
Take a~local parameter
$s$
at the point~
$R$.
It follows from (3) that
$\theta_1(K)=0$.
The following equality holds:
$$
\frac{d}{ds}\theta\biggl(\,\int\limits_{s(Q)}^{s(P)}s^*\omega+3K
\biggr)=
\theta_1\biggl(\,\int\limits_{s(Q)}^{s(P)}s^*\omega+3K\biggr)
\frac{\omega_1}{ds}+
\theta_2\biggl(\int\limits_{s(Q)}^{s(P)}s^*\omega+3K\biggr)
\frac{\omega_2}{ds}.
$$
Since
$\omega_2(R)=0$
and
$\theta_1(K)=0$,
we have
$$
\frac{d}{ds}\theta\biggl(\,\int\limits_{s(Q)}^{s(P)}s^*\omega+3K
\biggr)
=0
$$
for
$P=R$;
consequently, the function
$\theta(A(P)+K+2K)$
has zero of multiplicity~2 at~
$R$
or, equivalently, the function
$\theta(z+2K)$
has zero of multiplicity~2 on
$\Theta$
at~
$-K$.
Hence, the point
$-K$
is a~tangency point of~
$\Theta$
and
$\Gamma_1$.
Similarly, we can prove that
$\Theta$
and
$\Gamma_2$
are tangent at~
$K$.

Prove that
$\theta_{11}(-K)\ne 0$.
From (3) we obtain the identity
$$
\theta_{112}(z)\theta(z)-
\theta_{11}(z)\theta_{2}(z)-
2\theta_{12}(z)\theta_{1}(z)-
2c_3\theta_2(z)\theta(z)
$$
$$
=c_2\theta_2(z+2K)\theta(z-2K)+
c_2\theta(z+2K)\theta_2(z-2K)
$$
which implies that
$\theta_{11}(-K)\ne 0$;
otherwise we would have
$\theta_2(K)=0$,
but
$\Theta$
is a~smooth Riemann surface.
The lemma is proven.

From (3) we conclude that the following equality is valid
on
$\Gamma_1$
and
$\Gamma_2$:
$$
\theta_{11}(z)\theta(z)-
\theta_1^2(z)-c_3\theta^2(z)=0.
$$
Lemma~5 implies that the function
$\theta_1(z)$
on
$\Gamma_1$
and
$\Gamma_2$
has zeros of the first order at
$-K$
and
$K$;
consequently, we have the expansions
$$
\theta_2(z)=\theta_2(-K)+b_1\theta_1(z)+o(\theta_1(z)),
\quad
\theta_{11}(z)=\theta_{11}(-K)+d_1\theta_1(z)+o(\theta_1(z))
                            \eqno{(4)}
$$
on~
$\Gamma_1$
(in a~neighborhood of the point~$-K\in\Gamma_1$)
and the expansions
$$
\theta_2(z)=\theta_2(K)+b_2\theta_1(z)+o(\theta_1(z)),
\quad
\theta_{11}(z)=\theta_{11}(K)+d_2\theta_1(z)+o(\theta_1(z))
$$
on~
$\Gamma_2$
(in a~neighborhood of the point~$K\in\Gamma_2$),
where
$b_i,d_i\in{\Bbb C}$, $i=1,2$.

{\bf Lemma 6.} {\sl
The equalities
$b_1=b_2$ and $d_1=-d_2$
hold.
}

{\sc Proof.}
Denote by
$A_1$
the mapping
$\Gamma\rightarrow X$
which is defined by the formula
$A_1(P)=\int\nolimits_Q^P\omega-K,$
$P\in\Gamma$,
and suppose that
$A_2(P)=-A_1(P)$.
The range of
$A_1$
is the Riemann surface
$\Gamma_1$
and the range of~
$A_2$
is
$\Gamma_2$.
Since
$\theta(z)$
is an~even function,
$\theta_1(z)$
and
$\theta_2(z)$
are odd; consequently, from (4) we obtain
$$
\theta_2(A_2(P))=\theta_2(K)+b_1\theta_1(A_2(P))+\dots.
$$
Hence,
$b_1=b_2.$
Similarly, we can prove that
$d_1=-d_2.$
The lemma is proven.

Put
$b=b_1=b_2$ and $d=d_1=-d_2$.


Introduce the following functions in~
$M_c$:
$$
\psi_1=\frac{\theta(z+c+x)}{\theta(z)\theta(c-K+x)}
$$
$$
\times
\exp\left(-x_1\left(\partial_{z_1}\log\theta(z)-
\frac{b\theta_{11}(K)}{2\theta_2(K)}-\frac{d}{2}\right)
-x_2\partial_{z_2}\log\theta(z)\right),
                            \eqno{(5)}
$$
$$
\psi_2=\frac{\theta(z+c-2K+x)
\theta(z+2K)}{\theta^2(z)\theta(c-K+x)}
$$
$$
\times
\exp\left(-x_1\left(\partial_{z_1}\log\theta(z)+
\frac{b\theta_{11}(K)}{2\theta_2(K)}+\frac{d}{2}\right)
-x_2\partial_{z_2}\log\theta(z)\right).
\eqno{(6)}
$$
The functions
$\psi_1$ and $\psi_2$
constitute a~basis for the ${\cal D}$-module~
$M_c$~[3].

Denote by
$L^{ij}_{c,K}$ and $L^{ijk}_{c,K}$
the Nakayashiki operators
$L_{c,K}(\partial_{z_i}\partial_{z_j}\log\theta(z))$
and
$L_{c,K}(\partial_{z_i}\partial_{z_j}\partial_{z_k}\log\theta(z))$
in the basis
$\psi_1,\psi_2$
(the meaning of the subscript $K$ will be seen later). The operator
$\bigl[L_{c,K}^{ij}\bigr]_{11}$
looks like
$-\partial_{x_i}\partial_{x_j}+f_{c,K}^{ij}\partial_{x_1}+
g_{c,K}^{ij}\partial_{x_2}+h_{c,K}^{ij}$
and the operator
$[L_{c,K}(\partial_{z_i}\partial_{z_j}\log\theta(z))]_{12}$
is the operator of multiplication by some function
$H^{ij}_{c,K}(x),$
$i,j=1,2$
[3].

{\bf Lemma 7.} {\sl
The following equalities hold:
$$
 g^{11}_{c,K}=\frac{\theta_{11}(K)}{\theta_2(K)},
\quad
 g^{12}_{c,K}=\frac{b\theta_{11}(K)}{2\theta_2(K)}+
 \frac{\theta_{12}(K)}{\theta_2(K)}+\frac{d}{2},
\quad
 g^{22}_{c,K}=\frac{\theta_{22}(K)}{\theta_2(K)},
\quad
 f^{11}_{c,K}=0,
$$
$$
h^{11}_{c,K}=\partial^2_{x_1}\log\frac{\theta(c-3K+x)}{\theta(c-K+x)}
$$
$$
+\left(\partial_{x_1}\log\frac{\theta(c-3K+x)}{\theta(c-K+x)}+
\partial_{z_1}\log\theta(3K)+
\frac{b\theta_{11}(K)}{2\theta_2(K)}+\frac{d}{2}\right)^2
$$
$$
-\frac{\theta_{11}(K)}{\theta_2(K)}
\left(\partial_{x_2}\log\frac{\theta(c-3K+x)}{\theta(c-K+x)}+
\partial_{z_2}\log\theta(3K)\right)+c_3,
$$
$$
H^{11}_{c,K}=
\frac{2\theta_{11}(K)\theta(K+c+x)}
{\theta(3K)\theta(c-K+x)}
\exp(x_1(\frac{b\theta_{11}(K)}{\theta_2(K)}+d)),
$$
$$
H^{12}_{c,K}=\left(\frac{\theta_{12}(K)}{\theta_{11}(K)}+
 \frac{b}{2}+\frac{d\theta_2(K)}{2\theta_{11}(K)}\right)H^{11}_{c,K}-
 \frac{\theta_2(K)}{2\theta_{11}(K)}
 \partial_{x_1}H^{11}_{c,K},
$$
$$
H^{22}_{c,K}=\left(\frac{\theta_{22}(K)}{\theta_{11}(K)}+b+
\frac{d\theta_2(K)}{\theta_{11}(K)}\right)
H^{11}_{c,K}- \frac{\theta_2(K)}{\theta_{11}(K)}
 \partial_{x_2}H^{11}_{c,K}.
$$
}

{\sc Proof.}
Divide the equality
$$
-\partial_{x_i}\partial_{x_j}\psi_1+
f^{ij}_{c,K}\partial_{x_1}\psi_1+
g^{ij}_{c,K}\partial_{x_2}\psi_1+h^{ij}_{c,K}\psi_1+
H^{ij}_{c,K}\psi_2=
\partial_{z_i}\partial_{z_j}\log\theta(z)\psi_1
$$
by
$\exp(-x_1\partial_{z_1}\log\theta(z)-x_2\partial_{z_2}\log
\theta(z))$
and multiply by
$\theta^2(z)$.
Putting first
$z=-K$
and then
$z=K$,
we obtain
$g^{ij}_{c,K}$
and
$H^{ij}_{c,K}$.
Now, take
$z\in\Gamma_1$.
Using ~(4), we find that
$f^{11}_{c,K}=0$.
The following equality is valid for
$z\in\Gamma_1$:
$$
\partial^2_{x_1}\psi_1-
\frac{\theta_{11}(K)}{\theta_2(K)}\partial_{x_2}\psi_1-
h^{11}_{c,K}\psi_1+c_3\psi_1=0;
                            \eqno{(7)}
$$
consequently,
$$
h^{11}_{c,K}=
\partial^2_{x_1}\log\psi_1+
(\partial_{x_1}\log\psi_1)^2-
\frac{\theta_{11}(K)}{\theta_2(K)}\partial_{x_2}\log\psi_1
+c_3.
$$
Putting
$z=-3K\in\Gamma_1$,
we obtain
$h^{11}_{c,K}$.
The lemma is proven.

{\bf Lemma 8.} {\sl
The following equalities hold for the operator
$L_{c,K}=L_{c,K}(\lambda)$,
$\lambda\in{\cal A}_{\Theta}$:
$$
[L_{c,K}]_{21}=[L_{c-2K,-K}]_{12}\left(\frac{1}{c_2}
\bigl[L^{11}_{c,K}\bigr]_{11}-
\frac{c_3}{c_2}\right),
$$
$$
[L_{c,K}]_{22}=
[L_{c-2K,-K}]_{11}+
\frac{1}{c_2}[L_{c-2K,-K}]_{12}H^{11}_{c,K}.
$$
}

{\sc Proof.}
Replace
$K$
with
$-K$
and
$c$
with
$c-2K$
in the equality
$[L_{c,K}]_{11}\psi_1+[L_{c,K}]_{12}\psi_2=\lambda(z)\psi_1$
and multiply both sides by
$\frac{\theta(z+2K)}{\theta(z)}$.
Observe that $d$
goes into~
$-d$
and
$b$
remains the same upon this change.
We obtain
$$
[L_{c-2K,-K}]_{11}\psi_2+[L_{c-2K,-K}]_{12}
\frac{\theta(z+2K)\theta(z-2K)}{\theta^2(z)}
\psi_1=\lambda(z)\psi_2.
$$
Consequently,
$$
[L_{c-2K,-K}]_{12}\left(\frac{1}{c_2}
[L_{c,K}^{11}]_{11}-\frac{c_3}{c_2}\right)
\psi_1
$$
$$
+\left([L_{c-2K,-K}]_{11}+
\frac{1}{c_2}[L_{c-2K,-K}]_{12}H^{11}_{c,K}\right)\psi_2
=\lambda(z)\psi_2.
$$
The lemma is proven.

In particular, it follows from Lemma~8 that
$$
\bigl[L^{ij}_{c,K}\bigr]_{21}=H^{ij}_{c-2K,-K}
\left(\frac{1}{c_2}\bigl[L^{11}_{c,K}\bigr]_{11}-\frac{c_3}{c_2}
\right),
$$
$$
\bigl[L^{ij}_{c,K}\bigr]_{22}=
\bigl[L^{ij}_{c-2K,-K}\bigr]_{11}+
\frac{1}{c_2}H^{11}_{c,K}H^{ij}_{c-2K,-K}.
$$
Denote
the function
$H^{11}_{c-2K,-K}$ by ~$V$
and  denote
$H^{11}_{c,K}$ by ~$W$:
$$
V=\frac{2\theta_{11}(K)\theta(c-3K+x)}
{\theta(3K)\theta(c-K+x)}
\exp\left(-x_1\left(\frac{b\theta_{11}(K)}{\theta_2(K)}+d\right)
\right),
                                \eqno{(8)}
$$
$$
W=\frac{2\theta_{11}(K)\theta(K+c+x)}
{\theta(3K)\theta(c-K+x)}
\exp\left(x_1\left(\frac{b\theta_{11}(K)}{\theta_2(K)}+d\right)
\right).
                                \eqno{(9)}
$$
From Lemma~7 we obtain
$$
h_{c,K}^{11}=
\partial_{x_1}^2\log V+
\left(\partial_{x_1}\log V+
\frac{3b\theta_{11}(K)}{2\theta_2(K)}+
\frac{3d}{2}+
\partial_{z_1}\log\theta(3K)\right)^2
$$
$$
-\frac{\theta_{11}(K)}{\theta_2(K)}
(\partial_{x_2}\log V+
\partial_{z_2}\log\theta(3K))+c_3,
 \eqno{(10)}
$$
$$
h^{11}_{c-2K,-K}=
\partial_{x_1}^2\log W+
\left(\partial_{x_1}\log W-
\frac{3b\theta_{11}(K)}{2\theta_2(K)}-
\frac{3d}{2}-
\partial_{z_1}\log\theta(3K)\right)^2
$$
$$
+\frac{\theta_{11}(K)}{\theta_2(K)}
(\partial_{x_2}\log W-
\partial_{z_2}\log\theta(3K))+c_3.
\eqno{(11)}
$$
In a~neighborhood of
$z=-K$
on~
$\Gamma_1$
we have the expansion
$$
\frac{\theta_{12}(z)}{\theta(z)}=
\frac{a_2}{\theta_1^2(z)}+
\frac{a_1}{\theta_1(z)}+\dots,
\quad  a_1,a_2\in{\Bbb C},
                            \eqno{(12)}
$$
and in a~neighborhood of
$z=K$
on~
$\Gamma_2$,
the expansion
$$
\frac{\theta_{12}(z)}{\theta(z)}=
\frac{\tilde{a}_2}{\theta_1^2(z)}+
\frac{\tilde{a}_1}{\theta_1(z)}+\dots,
\quad
\tilde{a}_1,\tilde{a}_2\in{\Bbb C}.
$$
As in Lemma~6, we easily demonstrate that
$a_1=-\tilde{a}_1$.
Put
$a=a_1=-\tilde{a}_1$.

{\bf Lemma 9.} {\sl
The following hold:
$$
h^{12}_{c,K}=
\partial_{x_1}\partial_{x_2}\log V+
\left(\partial_{x_1}\log V+
\frac{3b\theta_{11}(K)}{2\theta_2(K)}+
\frac{3d}{2}+
\partial_{z_1}\log\theta(3K)\right)
$$
$$
\times(\partial_{x_2}\log V+
\partial_{z_2}\log\theta(3K))-
f^{12}_{c,K}\left(\partial_{x_1}\log V+
\frac{3b\theta_{11}(K)}{2\theta_2(K)}+
\frac{3d}{2}+
\partial_{z_1}\log\theta(3K)\right)
$$
$$
-
g^{12}_{c,K}(\partial_{x_2}\log V+
\partial_{z_2}\log\theta(3K))+
\partial_{z_1}\partial_{z_2}\log\theta(3K),
$$
$$
f^{12}_{c,K}=
\frac{\theta_2(K)}{2\theta_{11}(K)}
\biggl(h^{11}_{c,K}-c_3+
2\frac{d\theta_{12}(K)}{\theta_2(K)}-
\frac{2b\theta_{11}(K)\theta_{12}(K)}{\theta_2^2(K)}-
\frac{2a}{\theta_2(K)}
$$
$$
-\left(\frac{d}{2}-\frac{b\theta_{11}(K)}{2\theta_2(K)}\right)^2-
2\theta_{11}(K)e-\frac{\theta_{11}(K)}{\theta_2(K)}\alpha\biggr).
$$
}

{\sc Proof.}
In the proof of the lemma we suppose that
$z\in\Gamma_1$.
Then
$$
h^{12}_{c,K}=
\partial_{x_1}\partial_{x_2}\log\psi_1+
\partial_{x_1}\log\psi_1\partial_{x_2}\log\psi_1-
f^{12}_{c,K}\partial_{x_1}\log\psi_1
$$
$$
-
g^{12}_{c,K}\partial_{x_2}\log\psi_1+
\partial_{z_1}\partial_{z_2}\log\theta(z).
$$
 Putting
$z=-3K$,
we hence obtain
$h^{12}_{c,K}$.

For convenience, we denote
the local parameter
$\theta_1(z)$
on ~
$\Gamma_1$
at ~
$z=-K$ by $k^{-1}$.
From (3) and (4) we derive the expansion
$$
\frac{\theta_1(z)}{\theta(z)}=
\frac{\theta_{11}(z)}{\theta_1(z)}-
c_3\frac{\theta(z)}{\theta_1(z)}=
\theta_{11}(K)k+d+\frac{e}{k}+
o(k^{-1}),\quad e\in{\Bbb C}.
                            \eqno{(13)}
$$
Let
$$
\frac{\theta_2(z)}{\theta(z)}=
\gamma k^2+\beta k+\alpha+o(1),
\quad
\alpha,\beta,\gamma\in{\Bbb C}.
                            \eqno{(14)}
$$
Then
$\psi_1$
has the form
$$
\psi_1=\frac{1}{\theta(z)}
\left(1+\frac{\xi_1(x)}{k}+\frac{\xi_2(x)}{k^2}+
o\left(\frac{1}{k^2}\right)\right)\exp,
$$
$$
\exp=
\exp\biggl(-x_1\left(\left(\theta_{11}(K)k+d+\frac{e}{k}+
\dots\right)-
\frac{b\theta_{11}(K)}{2\theta_2(K)}-
\frac{d}{2}\right)
$$
$$
-x_2(\gamma k^2+\beta k+\alpha+\dots)\biggr).
$$
Equating  the coefficients of
$k^2\exp$, $k\exp$,
and
$\exp$
in~(7) to zero, we find that
$$
\gamma=-\theta_{11}(K)\theta_2(K),
\quad
\beta=b\theta_{11}(K)-d\theta_2(K)
$$
and
$$
h^{11}_{c,K}-c_3=-2\theta_{11}(K)
\partial_{x_1}\xi_1+
\left(\frac{d}{2}-\frac{b\theta_{11}(K)}{2\theta_2(K)}\right)^2+
2\theta_{11}(K)e+
\frac{\theta_{11}(K)}{\theta_2(K)}\alpha.
$$
Using ~(12)--(14), equate the coefficients of~
$k^2\exp$
and~
$k\exp$
in the identity
$$
-\partial_{x_1}\partial_{x_2}\psi_1+
f^{12}_{c,K}\partial_{x_1}\psi_1+
g^{12}_{c,K}\partial_{x_2}\psi_1+
h^{12}_{c,K}\psi_1=
\partial_{z_1}\partial_{z_2}\log\theta(z)\psi_1.
$$
We obtain
$a_2=\theta_{11}(K)\theta_2(K)$
and
$$
f^{12}_{c,K}=
-\theta_2(K)
\partial_{x_1}\xi_1+
\frac{d\theta_{12}(K)}{\theta_{11}(K)}-
\frac{b\theta_{12}(K)}{\theta_2(K)}-
\frac{a}{\theta_{11}(K)}.
$$
The lemma is proven.

{\bf Lemma 10.} {\sl
The coefficients of the $11$-components of the operators
$L_{c,K}({\cal A}_{\Theta})$
are rationally expressible in terms of ~
$V$
and its derivatives.
}

{\sc Proof.}
Take
$z\in\Gamma_1$.
Then
$\bigl[L_{c,K}^{12}\bigr]_{11}\psi_1=\partial_{z_1}\partial_{z_2}
\psi_1$.
Replace differentiation of
$\psi_1$
with respect to ~
$x_2$
on the left-hand side of this equality
with differentiation with respect to~
$x_1$
(by~(7)).
We obtain some third-order operator
$\widetilde{L}$
(in~$x_1$).
As follows from Lemmas~7 and~9, the  coefficients of ~$\widetilde{L}$
are rationally expressible in terms of ~
$V$
and its derivatives.

Let
$[L_{c,K}(\lambda)]_{11}$, $\lambda\in{\cal A}_{\Theta}$,
be an~arbitrary operator.
We have
$[L_{c,K}(\lambda)]_{11}\psi_1=\lambda\psi_1.$
As above, replace differentiation with respect to
$x_2$
with differentiation with respect to~
$x_1.$
We obtain some operator
$\widetilde{L}_1$
of order
$>3.$
The operators
$\widetilde{L}$
and
$\widetilde{L}_1$
commute
(as having a~family of common eigenfunctions
parametrized by the points of~
$\Gamma_1$).
 As demonstrated in~[9], the coefficients of
$\widetilde{L}_1$
are consequently polynomially expressible in terms of the
coefficients of
$\widetilde{L}$
and their derivatives. Hence, the coefficients of
$[L_{c,K}(\lambda)]_{11}$
are rationally expressible in terms of
$V$
and its derivatives. The lemma is proven.

Similarly, we can prove that the coefficients of the operators
$[L_{c-2K,-K}]_{11}$
are rationally expressible in terms of ~
$W$
and its derivatives.

Lemmas~7, 8, and~10 yield the following

{\bf Lemma 11.} {\sl
The coefficients of the second-order Nakayashiki operators
are rationally expressible in terms of
$W$, ~$V$,
and their derivatives.
}

The operator
$\bigl[L^{ijk}_{c,K}\bigr]_{11}$
has third order and the principal part
$\partial_{x_i}\partial_{x_j}\partial_{x_k}$,
whereas the operator
$\bigl[L^{ijk}_{c,K}\bigr]_{12}$
is of the first order~[3].

{\bf Lemma 12.} {\sl
The coefficients of the Nakayashiki operators of the third order
are rationally expressible in terms of ~
$W$, ~$V$,
and their derivatives.
}

{\sc Proof.}
Let
$[L_{c,K}]_{12}=u^1_{c,K}\partial_{x_1}+
u^2_{c,K}\partial_{x_2}+u^3_{c,K},$
where
$L_{c,K}$
is a~third-order operator. By Lemma~8, we have
$$
[L_{c,K}]_{21}=\bigl(u^1_{c-2K,-K}\partial_{x_1}+
u^2_{c-2K,-K}\partial_{x_2}+u^3_{c-2K,-K}\bigr)
$$
$$
\times\left(\frac{1}{c_2}\left(-\partial^2_{x_1}+
\frac{\theta_{11}(K)}{\theta_2(K)}\partial_{x_2}+h^{11}-c_3\right)
\right),
$$
$$
[L_{c,K}]_{22}=[L_{c-2K,-K}]_{11}+\frac{1}{c_2}
\bigl(u^1_{c-2K,-K}\partial_{x_1}+
u^2_{c-2K,-K}\partial_{x_2}+u^3_{c-2K,-K}\bigr)W.
$$
By Lemma~10, the coefficients of
$[L_{c,K}]_{11}$
and
$[L_{c-2K,-K}]_{11}$
are rationally expressible in terms of ~
$W$, ~$V$,
and their derivatives; hence, the coefficient of~
$\partial^3_{x_2}$
in the 21-component of the commutator
$[L^{11}_{c,K},L_{c,K}]=0$
has the form
$$
-\frac{2\theta_{11}^2(K)}{c_2\theta_2^2(K)}u^2_{c-2K,-K}+F^2=0,
$$
where the function
$F^2$
is rationally expressible in terms of
$W$, ~$V$,
and their derivatives.
The coefficients of
$\partial_{x_1}\partial^2_{x_2}$
and
$\partial^2_{x_2}$
in this component have the form
$$
-\frac{2\theta_{11}^2(K)}{c_2\theta_2^2(K)}u^i_{c-2K,-K}+F^i=0,
\quad
i=1,3,
$$
where the functions
$F^i$
are rationally expressible in terms of
$V$, $W$, $u^2_{c-2K,-K}$,
and their derivatives. Consequently,
$u^i_{c-2K,-K}$ and $u^i_{c,K}$
are rationally expressible in terms of
$V$, ~$W$,
and their derivatives.
The lemma is proven.

The second- and third-order operators generate the whole ring
$L_{c,K}({\cal A}_{\Theta})$~
[3];
thereby Lemmas~11 and~12 yield Theorem~3.

Write down formulas for the constants
$c_j$
(from the introduction).

Multiply both sides of~(3) by
$\theta^2(z)$
and differentiate with respect to~
$z_2$.
Putting $z=K$,
we obtain~
$c_2$.
Putting
$z=3K$ in~(3), we find
$c_3$.
From Lemmas~7 and~9 and formulas (10) and~(11) we find the
following formulas for the other constants~
$c_j$:
$$
c_1=\frac{\theta_{11}(K)}{\theta_2(K)},\quad
c_2=-\frac{\theta_{11}(K)}{\theta(3K)},\quad
c_3=\frac{\theta_{11}(3K)}{\theta(3K)}-
\frac{\theta_1^2(3K)}{\theta^2(3K)},
                            \eqno{(15)}
$$
$$
c_4=\frac{3}{2}bc_1+\frac{3d}{2}+
\partial_{z_1}\log\theta(3K),
\quad
c_5=\partial_{z_2}\log\theta(3K),
                            \eqno{(16)}
$$
$$
c_6=\frac{1}{2}bc_1+\frac{d}{2}+
 \frac{\theta_{12}(K)}{\theta_2(K)},
\quad
c_7=-c_5c_6+\partial_{z_1}\partial_{z_2}\log\theta(3K),
                                \eqno{(17)}
$$
$$
c_8=
2\frac{d\theta_{12}(K)}{\theta_2(K)}-
\frac{2b\theta_{11}(K)\theta_{12}(K)}{\theta_2^2(K)}-
\frac{2a}{\theta_2(K)}
$$
$$
-
\left(\frac{d}{2}-\frac{b\theta_{11}(K)}{2\theta_2(K)}\right)^2-
2\theta_{11}(K)e-
\frac{\theta_{11}(K)}{\theta_2(K)}\alpha-c_3.
\eqno{(18)}
$$
The constants
$a$, $b$, $d$, $e$,
and
$\alpha$
are determined from ~
(12), (4), (13), and (14).

{\bf References.}

[1] Nakayashiki~A.
      Structure of Baker--Akhiezer modules of
principally polarized abelian varieties, commuting partial
differential operators and associated integrable systems.
       Duke Math.~J., vol. 62, N. 2, 315--358 (1991).

[2] Nakayashiki~A. Commuting partial differential operators
and vector  bundles over abelian  varieties.
      Amer.~J. Math., vol. 116, 65--100 (1994).

[3] Mironov A.~E. Commutative rings of~differential operators
connected with~two-dimensional abelian  varieties. Sibirsk. Mat. Zh.,
     vol. 41, N. 6, 1389--1403 (2000).

[4]
Feldman J., Knorrer~H., and  Trubowitz~E.
There is no two-dimensional analogue of Lame's equation.
       Math. Ann.,
     vol.   294,
      N.     2,       295--324 (1992).

[5]
Dubrovin~B.~A. and Natanzon~S.~M.
      Real two-gap solutions of the sine-Gordon equation.
       Funktsional. Anal. i Prilozhen.,
      vol.   16,
      N.   1, 27--43 (1982).

[6]
 Dubrovin~B.~A. and  Novikov~S.~P.
      Ground states in a periodic field. Magnetic Bloch functions and
       vector
      bundles.
       Dokl. Akad. Nauk SSSR,
     vol.   253, N.   6,   1293--1297 (1980).

[7]
 Mumford~D.       Tata Lectures on Theta,
Tata Lectures on Theta I, II,
Birkh\"auser, Boston-Basel-Stuttgart,
1983, 1984.

[8]
 Griffiths~F. and Harris~J.,
       Principles of Algebraic Geometry,
   N.Y. - Chichester - Brisbane - Toronto :
   John Wiley \& Sons, 1978.

[9]
 Krichever~I.~M.
       The methods of algebraical geometry
      in the theory of nonlinear equations.
      Uspekhi Mat. Nauk, vol. 32, N. 6, 183--208 (1997)

[10]
 Fay~J.~D., Theta functions on Riemann surfaces.
Lecture Notes in Math.; {\bf 352},
Springer-Verlag, Berlin, Heidelberg, and  New York (1973).

[11]
Ta\u\i manov~I.~A.
      Secants of abelian varieties, theta-functions,
      and soliton equations. Uspekhi Mat. Nauk.
      vol.    52, N.  1,    149--224 (1997).

 \vskip3mm

 {\sl
 Sobolev Institute of Mathematics,
 Novosibirsk}

 mironov@math.nsc.ru

\enddocument